\documentclass[aip,
 amsmath,amssymb,
preprint,%
]{revtex4-1}

\usepackage{graphicx}
\usepackage{dcolumn}
\usepackage{bm}
\usepackage[mathlines]{lineno}

\usepackage[utf8]{inputenc}
\usepackage[T1]{fontenc}
\usepackage{mathptmx}
\usepackage{etoolbox}

\usepackage{amsmath}
\usepackage{color}
\usepackage[caption=false]{subfig}
\usepackage{siunitx}
\usepackage[version=4]{mhchem}

\makeatletter
\def\@email#1#2{%
 \endgroup
 \patchcmd{\titleblock@produce}
  {\frontmatter@RRAPformat}
  {\frontmatter@RRAPformat{\produce@RRAP{*#1\href{mailto:#2}{#2}}}\frontmatter@RRAPformat}
  {}{}
}%
\makeatother
\begin{document}

\preprint{AIP/123-QED}

\title[Vapor compression and energy dissipation in a collapsing laser-induced bubble]{Vapor compression and energy dissipation in a collapsing laser-induced bubble}
\author{D. B. Preso}
\affiliation{Institute of Mechanical Engineering, École Polytechnique Fédérale de Lausanne, Avenue de Cour 33 Bis, 1007 Lausanne, Switzerland}
\email{davide.preso@epfl.ch}

\author{D. Fuster}%
\affiliation{Sorbonne Université, Centre National de la Recherche Scientifique, UMR 7190, Institut Jean Le Rond $\partial$’Alembert, F-75005 Paris, France}

\author{A. B. Sieber}
\affiliation{Institute of Mechanical Engineering, École Polytechnique Fédérale de Lausanne, Avenue de Cour 33 Bis, 1007 Lausanne, Switzerland}

\author{D. Obreschkow}
\affiliation{International Centre for Radio Astronomy Research (ICRAR), University of Western Australia, Crawley, WA 6009, Australia}

\author{M. Farhat}
\affiliation{Institute of Mechanical Engineering, École Polytechnique Fédérale de Lausanne, Avenue de Cour 33 Bis, 1007 Lausanne, Switzerland}

\date{\today}

\begin{abstract}
The composition of the gaseous phase of cavitation bubbles and its role on the collapse remains to date poorly understood. In this work, experiments of single cavitation bubbles in aqueous ammonia serve as a novel approach to investigate the effect of the vapor contained in a bubble on its collapse. We find that the higher vapor pressure of more concentrated aqueous ammonia acts as a resistance to the collapse, reducing the total energy dissipation. In line with visual observation, acoustic measurements, and luminescence recordings, it is also observed that higher vapor pressures contribute to a more spherical collapse, likely hindering the growth of interface instabilities by decreasing the collapse velocities and accelerations.
Remarkably, we evidence a strong difference between the effective damping and the energy of the shock emission, suggesting that the latter is not the dominant dissipation mechanism at collapse as predicted from classical correction models accounting for slightly compressible liquids.
Furthermore, our results suggest that the vapor inside collapsing bubbles gets compressed, consistently with previous studies performed in the context of single bubble sonoluminescence, addressing the question about the ability of vapors to readily condense during a bubble collapse in similar regimes.
These findings provide insights into the identification of the influence of the bubble content and the energy exchanges of the bubble with its surrounding media, eventually paving the way to a more efficient use of cavitation in engineering and biomedical applications.
\end{abstract}

\maketitle

\section{Introduction}
\label{sec:introduction}
\par The collapse of cavitation bubbles often leads to shock waves emission, light radiation, and rebound bubbles.
The occurrence of these phenomena suggests the presence of a gaseous phase within the bubble, which is highly compressed during the collapse.
However, its nature and influence on the bubble dynamics is to date still a subject of debate.
Furthermore, although it is widely accepted that the non-condensable gas within the bubble undergo adiabatic compression, the role of condensable vapors remains vague as equilibrium conditions are at best only satisfied at the bubble interface but not in the bubble interior. \cite{StoreySzeri,Hauke,FusterHaukeDopazo}
This problem was already introduced in the last century by \citeauthor{Plesset} \cite{Plesset}, who speculated about the inability of vapor to change phase at the same rate of the bubble shrinkage.
Successively, several sophisticated numerical models have been proposed to capture the influence of phase change on the bubble dynamics.
\citeauthor{Fujikawa} \cite{Fujikawa}, and later \citeauthor{Akhatov1} \cite{Akhatov1}, developed numerical models which include liquid compressibility, and heat and mass transfer.
They investigated the incidence of non-equilibrium processes at the bubble wall due to thermal inertia of condensing vapor, concluding that this could lead to the occurrence of supercritical conditions at the final stage of the collapse.
They therefore highlighted the possibility of the vapor to behave as a non-condensable gas in the case where the volume reduction rate of the bubble was much higher than the condensation rate.
In addition, \citeauthor{Akhatov1} \cite{Akhatov1} introduced a sticking coefficient of water vapor, which played as a tuning parameter to fit experimental data and predicted the condensed vapor at the bubble-liquid interface.
Their work was further endorsed by \citeauthor{Szeri} \cite{Szeri}, who studied the heat and mass transfer during cavitation bubbles collapse, concluding that the latter occurs so fast that thermal diffusion and phase change effects are nearly obviated, as the vapor condensation rate is much slower than the bubble volume reduction rate.
More recently, \citeauthor{Magaletti} \cite{Magaletti} conducted a similar numerical investigation considering phase change, occurrence of supercritical conditions, thermal conduction, and liquid compressibility effects, reporting the disappearance and reappearance of liquid-vapor interface during the final stage of the collapse because of a transition to super-critical conditions of the vapor.
They also concluded that, in agreement with \citeauthor{Fujikawa} \cite{Fujikawa}, purely vapor bubble may be able to emit shock waves at collapse. Lately, \citeauthor{Liang} \cite{Liang} studied the transition from nonlinear to linear oscillations of collapsing cavitation bubbles.
They developed a novel approach based on the \citeauthor{Gilmore} \cite{Gilmore} model, with which they could fit the progressive condensation of water vapor during nonlinear bubble oscillations from experimental data by means of a tuning parameter, obtaining in turn the partial pressure of condensable vapor and non-condensable gas within the bubble.
The results were in good agreement with the ones of \citeauthor{Akhatov1} \cite{Akhatov1}.
Following this approach, \citeauthor{Wen} \cite{Wen} were able to track the bubble dynamics of millimeter-sized spherical cavitation bubbles up to the fourth oscillation.
The relevance of phase change on the bubble motion has been also discussed theoretically. Already for linear oscillations it is possible to distinguish regimes where the vapor is trapped inside the bubble while keeping equilibrium conditions at the interface \cite{FusterMontel,Bergamasco}.
For strongly non-linear oscillations, \citeauthor{FusterHaukeDopazo} \cite{FusterHaukeDopazo} have shown that this asymptotic limit is reached for large values of the accommodation coefficient, where the net flux across the interface is eventually dominated by diffusion effects \cite{Szeri}. 
Thus, the relevance of non-equilibrium conditions at the interface and the consequences of it on the bubble motion remain an open problem that needs of careful experimental investigations.
\par Irrespective from the fact that equilibrium conditions are sustained at the interface or not, the influence of phase change on the dynamics of bubbles has been experimentally confirmed in several works.
In relation to single bubble sonoluminescence, \citeauthor{VazquezPutterman} \cite{VazquezPutterman} observed an increased collapse cushioning and decreased light emission at increasing water temperature.
The latter observation was further confirmed by \citeauthor{Toegel} \cite{Toegel} and \citeauthor{Hopkins} \cite{Hopkins}, who highlighted the importance of the partial pressure of vapor trapped within the bubble during the collapse on the intensity of light emission.
Later, \citeauthor{TinguelyThesis} \cite{TinguelyThesis} and \citeauthor{Phan} \cite{Phan} investigated the dynamics of laser-induced cavitation bubbles in water at different temperatures, showing that the higher the water temperature, hence the vapor pressure, the larger the rebound bubble. 
To explain these effects, numerical models based on 
the slightly compressible versions of the Rayleigh-Plesset equation proposed by \citeauthor{KellerMiksis}\cite{KellerMiksis} and \citeauthor{Gilmore} \cite{Gilmore}, use the bubble internal pressure to fit the bubble radius evolution \cite{Liang,SupponenLuminescence,Zhang}. However, due to technological challenges involved in measuring the bubble contents at the sub-millimeter and sub-millisecond scale within the bubbles, direct probing of the inner bubble pressure remains uncertain \cite{Liu} and may hinder some limitations of the model when using experimental data to fit the evolution of the bubble radius. Among others, some limitations of these models are that numerically describe the dynamics of perfectly-spherical cavitation bubbles neglecting effects such as chemical reactions, phase change or strongly non-linear effects related to liquid and gas compressibility. In addition to the aforementioned uncertainties
regarding the modelling of phase change processes,
the influence of non-spherical deformations deserves particular attention as its effect has been indeed observed in various studies.
\citeauthor{Brennen} \cite{Brennen} investigated the bubble fission process due to bubble shape instabilities, concluding that the energy dissipated by the mixing and turbulence due to bubble fission may be preponderant compared to the conventional viscous and acoustic damping.
\citeauthor{Delale} \cite{Delale} developed a numerical model also accounting for deviations from sphericity, confirming the results of \citeauthor{Brennen} \cite{Brennen}.
Moreover, \citeauthor{SupponenJets} \cite{SupponenJets} experimentally investigated deformed cavitation bubbles and reported that bubbles experience weak jetting phenomena even with reduced anisotropy.
Bubble shape perturbations have been also largely investigated in relation to single bubble sonoluminescence,
where the development of hydrodynamic instabilities at the bubble-liquid interface has been shown to determine the stability diagrams in which light emission is observed \cite{BrennerHilgenfeldtLohse}.
\par In this work we investigate the influence of the vapor content during the collapse of laser-generated single cavitation bubbles in aqueous ammonia by systematically varying the ammonia mass fraction $w_{\ce{NH3}}$ in solution. The latter two-component solutions have similar densities, but different $p^*_\text{v}$ values, which we use to investigate the influence of the bubble internal composition.
Compared to single bubble sonoluminescence experiments, the main difference is that it is possible to investigate the influence of the vapor content in regimes in conditions where the bubble oscillation is not stable gaining further insights about
the role of phase change on extremely violent transient collapses.
In addition, because the system's temperature is kept constant, we avoid some problems related to thermal expansion-related misalignment of optical components, sensitivity variation of measuring instruments, and change of laser energy absorbed by the liquid at bubble generation.
These experiments allow us to investigate the influence of vapor content on various variables including (i) the rebound size and collapse time, (ii) microscopic shape of rebound bubble, (iii) luminescence, (iv) radiated shock at collapse, and (v) liquid pressure build up prior to final collapse. 
Our findings show that all five vary significantly with the \ce{NH3} concentration in solution, evidencing the significant role of the latter on the bubble collapse and supporting the notion of vapor compression.
Furthermore, our measurements provide evidence of the effect of the bubble contents on the acoustic emission at bubble collapse.

\section {Materials and methods}
\subsection{Laser-induced cavitation bubbles}
\label{sec:methods}
\begin{figure}
    \centering
    \includegraphics[width=0.7\textwidth]{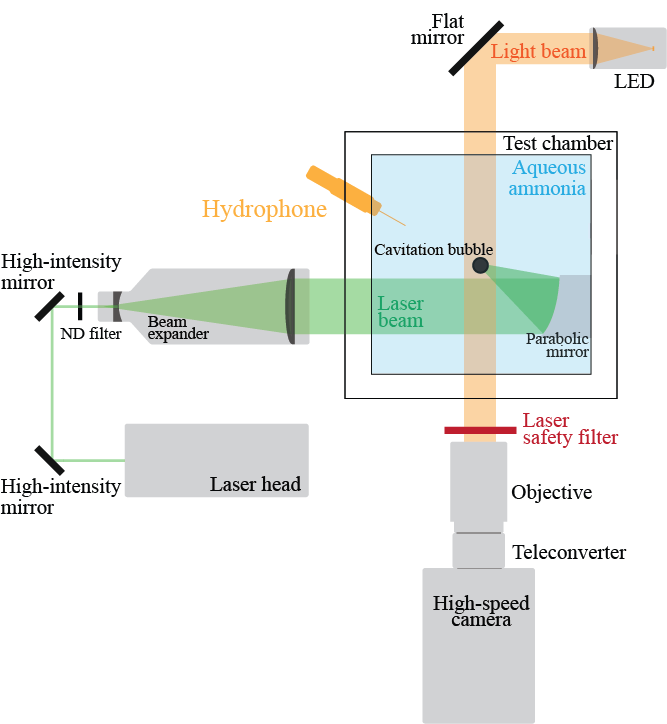}
    \caption{Top-view schematic of the experimental apparatus.}
    \label{fig:experimentalSetup}
\end{figure}

The experimental apparatus relies on a laser-based technique for the generation of single cavitation bubbles \cite{Obreschkow,SupponenExperiment,Sieber1,Sieber2} schematically shown in Figure \ref{fig:experimentalSetup}.
A bubble arises from the plasma generated by a 9-ns Nd:YAG laser pulse (Quantel CFR 400, 532 nm) focused into a point at the centre of an extended volume of liquid (aqueous ammonia in this work).
The laser beam, redirected with a set of high-intensity mirrors, is enlarged tenfold with a beam expander and focused with an off-axis parabolic mirror, immersed into the liquid in the test chamber.
The anisotropy parameter $\zeta$ for the generated bubbles was kept below the topological limit between spherical and toroidal collapse ($\zeta < 4 \times 10^{-4}$), such that any re-entrant jet did not pierce the bubble at collapse (a detailed description of $\zeta$ is given by \citeauthor{Obreschkow2} \cite{Obreschkow2} and \citeauthor{SupponenJets} \cite{SupponenJets}).
To this end, we generated bubbles with a maximum radius $R_0 \approx 1.5$ mm.
This size was achieved by adjusting the laser beam energy with a neutral-density filter.
\begin{figure*}
    \centering
    \includegraphics[width=0.96\textwidth]{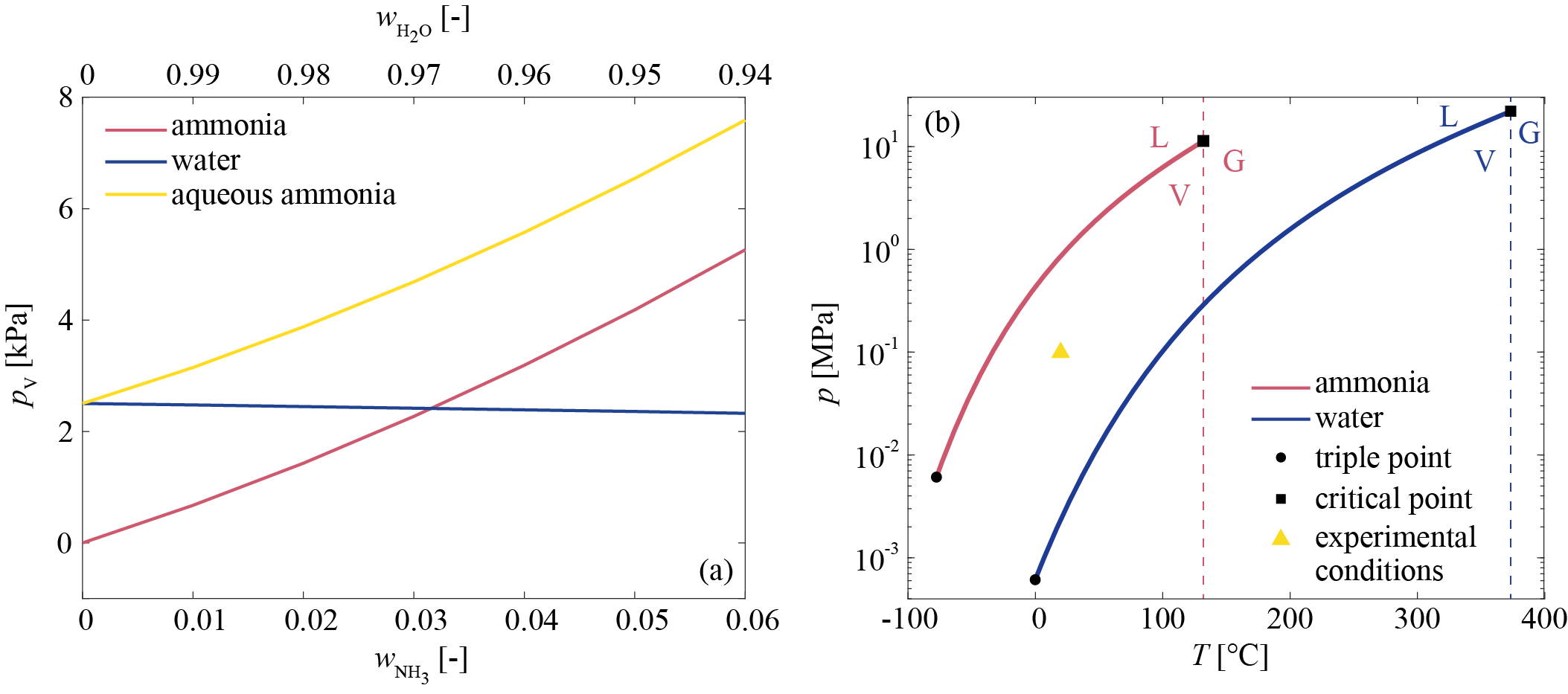}
    \caption{(a) Total vapor pressure of aqueous ammonia solutions as a function of the mass fraction of ammonia $w_{\text{NH}_3}$ (and relative water mass fraction $w_{\text{H}_2\text{O}}$) at \SI{21}{\celsius} from \citeauthor{Perry} \cite{Perry}. The plot also displays the partial vapor pressure of the solution's components.
    (b) Phase diagram of pure water and pure ammonia from \citeauthor{Perry} \cite{Perry}. The solid lines show the liquid-vapor phase boundary for both substances. The black circles indicate the triple point, whereas the black squares indicate the critical point. L, V, and G stands for liquid, vapor, and gas, respectively. The triangle indicates the experimental conditions.}
    \label{fig:phaseDiagram}
\end{figure*}
Aqueous ammonia was contained in a transparent gas-tight box (cubic shape, 18 cm edge length), which prevented ammonia leaks and hence concentration variations.
A total of six ammonia mass fractions $w_{\text{NH}_3}$ were exploited in the experiments, ranging from 0 (pure water) to 0.05.
The pressure $p_\infty$ and the temperature $T_\infty$ of the liquid at rest were kept constant and equal to the atmospheric pressure ($\approx 97$ kPa) and ambient temperature ($\approx \SI{20}{\celsius}$), respectively.
\par A high-speed camera (Shimadzu HPV-X2) filming at up to 10 million frames per second, backlighted by a collimated LED, recorded shadowgrams of the bubble.
The camera is equipped with a 105 mm objective (Nikon AF-S Micro 1:2.8 GED), and a 2x teleconverter (Nikon AF-S  TC-20E).
A safety filter ($532$ nm high-pass filter) is mounted in front of the objective to prevent reflected laser beams to accidentally reach the camera sensor.
The camera records 256 frames per film at a fixed resolution of 400x250 pixels at all frame rates below 10 million. 
At the latter, the resolution is halved.
The results of bubble dynamics presented in this work were obtained from high-speed recordings at 500’000 fps, whereas luminescence was recorded at 10 million fps.
The instantaneous radius $R$ of the bubble was obtained with automated image processing from high-speed recordings by retrieving the equivalent radius $R_\text{eq}$ of the bubble axial cross sectional area $A$ assuming spherical symmetry: $R = R_\text{eq} = \sqrt{A/\pi}$.
The bubble cross sectional area is represented by the black area in the bubble shadowgram recorded by the camera.
\par A needle-hydrophone placed perpendicularly to the bubble walls and 32.9 mm away from its center recorded the shock waves generated upon bubble generation and collapse.
At the instant of bubble maximum expansion, the potential energy of the bubble $E_\text{p0}$ can be written as
\begin{equation}
    E_\text{p0} = V_0 \left(p_\infty-p_\text{v}\right),
\end{equation}
where $V_0$ is the volume of the bubble at maximum expansion, $p_\text{v}$ is the vapor partial pressure within the bubble, conventionally equal to $p^*_\text{v}(T_\infty)$, and $p_\infty$ is the liquid pressure in the far field equal to the atmospheric pressure.
Accordingly, the potential energy of the rebound bubble at the instant of maximum expansion is $E_\text{p1} = V_1 \left(p_\infty-p_\text{v}\right)$, where $V_1$ is the maximum volume of the first rebound bubble.
Potential energy loss $\left(E_\text{p1} < E_\text{p0}\right)$ is a consequence of dissipation mechanisms at collapse, including the emission of shock waves away from the bubble. 
These effects dampen the bubble oscillations, progressively reducing the potential energy of the system, thereby diminishing the amplitude of succeeding rebounds.
The total energy loss $D_\text{tot}$ during the first collapse can be expressed as the difference in potential energy of the bubble between the moments of maximum expansion in the first oscillation and in the first rebound as
\begin{equation}
    D_\text{tot} = \left(V_0-V_1\right) \left(p_\infty - p_\text{v}\right).
    \label{eq:energybalance}
\end{equation}
The normalized value of the total energy loss is defined as $\widehat D_\text{tot} = D_\text{tot}/E_\text{p0}$.
On the other hand, assuming spherical propagation, the energy of the emitted shock waves $E_\text{sw}$ is calculated from far-field hydrophone measurements as \cite{SupponenShockwaves}
\begin{equation}
    E_\text{sw} = \frac{4\pi l^2}{G \rho c} U_\text{hyd,max}^b  \int U_\text{hyd}(t)^2 dt,
\end{equation}
where $l$ is the distance between the center of the bubble and the hydrophone sensor, $G$ is a calibration constant, $\rho$ is the density of the liquid, $c$ is the speed of sound in the medium, $U_\text{hyd}(t)$ is the signal of the hydrophone over time $t$, $U_\text{hyd,max}$ is the maximum value of $U_\text{hyd}(t)$, and $b$ is a factor which takes into account nonlinear dissipation effects during the shock wave propagation.
\citeauthor{VogelBuschParlitz} \cite{VogelBuschParlitz} showed that acoustic energy may be largely underestimated if determined from far-field measurements only.
In reality, shock waves dissipation and spreading of the shock width are nonlinear. 
Shock wave pressure decays proportionally to $r^{-1.1}$ even in the far field, with very fast decay close to the emission center proportional to $r^{-2}$, where $r$ is the radial coordinate with origin at the shock center \cite{VogelBuschParlitz,SupponenShockwaves,Doukas,Schoeffmann,VogelLauterbornTimm,DennerSchenke,Brujan}.
The use of $U_\text{hyd,max}^b$ to correct far-field hydrophone measurements was proposed by \citeauthor{SupponenShockwaves} \cite{SupponenShockwaves}
They found only slight nonlinear dissipation effects and computed $b \approx 0.45$.
In this work, the calibration constant $G$ for conversion from into pressure units is implied, so the shock wave energy $E_\text{sw}$ is reported with arbitrary units.
Hereinafter, we call $E_\text{sw}^\text{gen}$ the energy of the shock waves emitted at bubble generation, and $E_\text{sw}^\text{coll}$ those emitted at collapse.

\subsection{Aqueous ammonia solutions}
Aqueous ammonia solutions were prepared from distilled water and commercial aqueous ammonia (VWR, 25\% (w/w) ammonia content) by injecting them directly into the test chamber.
When dissolved in water, ammonia weakly dissociates as follows:
\begin{equation}
    \begin{split}
        \ce{NH3_{(g)}} & \rightleftharpoons \ce{NH3_{(aq)}},\\
        \ce{NH3_{(aq)}} + \ce{H2O_{(l)}} & \rightleftharpoons \ce{NH4+_{(aq)}} + \ce{OH^-_{(aq)}}.
    \end{split}
\end{equation}
Owing to the highly volatile nature of aqueous ammonia, we used extreme caution during the injection phase, and throughout the experimental campaign to prevent gas leaks, hence concentration variations.
\par Vapor pressure data of aqueous ammonia solutions are reported in Figure \ref{fig:phaseDiagram}(a).
The density $\rho$ of aqueous ammonia with a mass fraction from 0 to 0.05 is taken from \citeauthor{Perry} \cite{Perry} as 998.2, 993.9, 989.5, 985.3, 981.1, and 977.0 kg m$^{-3}$, respectively. 
The speed of sound $c$ of aqueous ammonia is retrieved from high-speed movies and is 1479, 1487, 1494, 1510, 1516, and 1535 m s$^{-1}$ for the same solutions cited before, respectively.
The critical point of ammonia, as reported in Figure \ref{fig:phaseDiagram}(b), is defined at \SI{132.5}{\celsius} and \SI{11.28}{MPa}, whereas the critical point of water is defined at \SI{374}{\celsius} and \SI{22.06}{MPa} \cite{Perry}.
At experimental conditions, ammonia and water coexists with the liquid phase in vapor form and can therefore be liquefied by compression only.
As reported by \citeauthor{Narita} \cite{Narita}, who investigated the dissolved oxygen in aqueous ammonia, the concentration of dissolved gas decreases with increasing ammonia concentration. 

\subsection{Analytical models for bubble motion}
For a bubble in an incompressible liquid, the bubble dynamics can be described by the Rayleigh-\citeauthor{Plesset} \cite{Plesset} equation
\begin{equation}
    R \ddot R + \frac{3}{2} \dot R^2 = \frac{p_\text{I} - p_\infty}{\rho},
\end{equation}
where the liquid pressure at the bubble-liquid interface $p_\text{I}$ is defined as
\begin{equation}
    p_\text{I} = p_\text{v}(T_\infty) + p_\text{g0} \left(\frac{R_0}{R}\right)^{3\gamma} - \frac{2 S}{R} - 4 \mu \frac{\Dot{R}}{R}.
\end{equation}
Here, $p_\text{g0}$ is the partial pressure of the non-condensable gas within the bubble, $\gamma$ is the heat capacity ratio of the gaseous phase being compressed, and $S$ and $\mu$ are the surface tension and the dynamic viscosity of the liquid, respectively.
The dotting indicates derivation in time.
However, this model is not able to reproduce experimental results, implying that liquid viscosity only cannot explain the damping experimentally observed during the collapse.
The Rayleigh-\citeauthor{Plesset}  \cite{Plesset} model is therefore not suited for fitting the bubble radius from experimental data.
As an alternative, the \citeauthor{KellerMiksis} \cite{KellerMiksis} and the \citeauthor{Gilmore} \cite{Gilmore} models introduce slightly compressibility effects in the liquid which allow to reproduce the bubble dynamics until the first rebound.
The classical \citeauthor{KellerMiksis} \cite{KellerMiksis} model reads
\begin{eqnarray}
    \left(1-\mathit{Ma}\right)R\Ddot{R} &+& \frac{3}{2}\left(1-\frac{\mathit{Ma}}{3}\right)\Dot{R}^2 = \\ && \frac{1}{\rho} \left(1+\mathit{Ma}\right) \left(p_\text{I}-p_\infty\right) + \frac{R}{\rho c} \frac{\partial p_\text{I}}{\partial t},
\end{eqnarray}
where $\mathit{Ma} = \dot{R} c^{-1}$ is the Mach number.
The \citeauthor{Gilmore} \cite{Gilmore} model reads
\begin{eqnarray}
    \left(1-\mathit{Ma}\right)R\Ddot{R} &+& \frac{3}{2}\left(1-\frac{\mathit{Ma}}{3}\right)\Dot{R}^2 =\\ && \left(1+\mathit{Ma}\right) H + \frac{R}{c} \left(1-\mathit{Ma}\right) \frac{\partial H}{\partial t},
\end{eqnarray}
where $H$ is the enthalpy difference between the pressure at the bubble-liquid interface and the liquid pressure in the far field, defined as
\begin{equation}
    H = \int_{p_\infty}^{p_\text{I}} \frac{dp}{\rho},
\end{equation}
where $\rho$ is treated as a function of the local pressure of the liquid $p$.
In both the \citeauthor{KellerMiksis} \cite{KellerMiksis} and \citeauthor{Gilmore} \cite{Gilmore} models, liquid compressibility acts as the preponderant mechanism controlling the bubble damping.
For sufficiently intense collapses, surface tension and viscosity play only a negligible role. \cite{PlessetProsperetti,LauterbornKurz}
The contribution of surface tension and viscosity to the bubble wall pressure scales with $1/R$, thus minimally affecting the dynamics of millimeter-sized bubbles as presented in this work. \cite{LauterbornKurz} 
Correspondingly,
\citeauthor{Liang} \cite{Liang} reported a significant effect of surface tension and viscosity in micrometer-sized bubbles.
\par A detailed procedure for fitting the bubble dynamics from experimental data is described in Appendix \ref{appAnalyticalModels}.

\section{Results and discussion}
\label{sec:results}
\subsection{Bubble dynamics in aqueous ammonia}
\begin{figure*}
    \centering
    \includegraphics[width=0.75\textwidth]{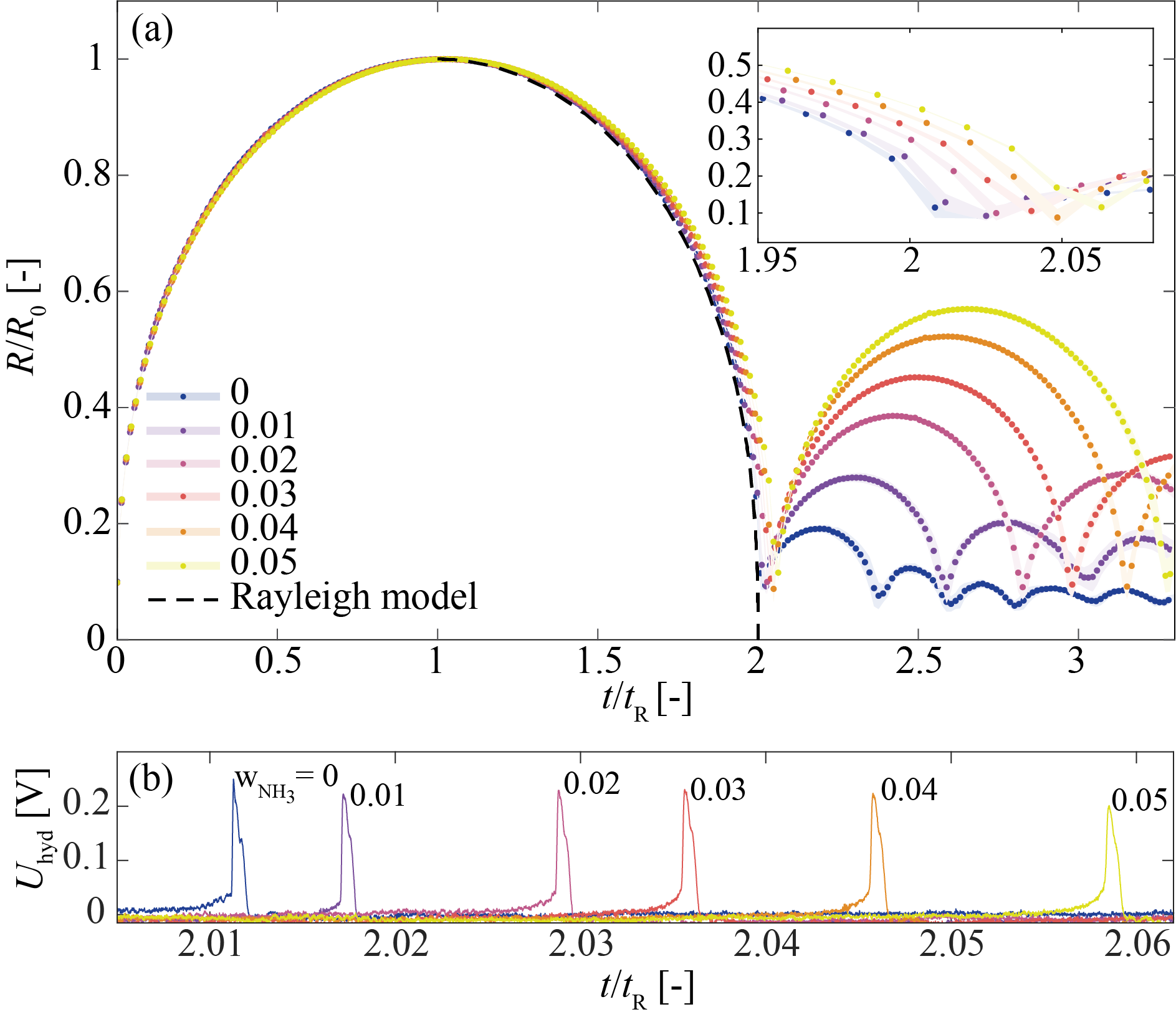}
    \caption{(a) Radial evolution over time in normalized coordinates of single laser-induced cavitation bubbles generated in six different aqueous ammonia solutions ($w_{\text{NH}_3}$ from 0 to 0.05). The points are discrete data averaged over 10 different experimental measurements. The color code indicates $w_{\text{NH}_3}$. The shaded area shows the standard deviation of the experimental measurements. The black dashed line is the solution of the \citeauthor{Rayleigh} \cite{Rayleigh} model for an empty collapsing bubble. 
    The nested plot is a magnification of the final stage of the collapse.
    (b) Raw hydrophone signal from acoustic measurements over time in normalized coordinate of bubbles collapsing in six different aqueous ammonia.
    The time $t=0$ is set at the instant when the generation shock is recorded.}
    \label{fig:rt}
\end{figure*}

Figure \ref{fig:rt}(a) shows the radial evolution in time in normalized coordinates, $R/R_0$ versus $t/t_\text{R}$, of single cavitation bubbles in aqueous ammonia, with $R_0$ the radius of the bubble at its maximum expansion, and $t_\text{R}=0.915 R_0 \sqrt{\rho/p_\infty}$ the \citeauthor{Rayleigh} \cite{Rayleigh} collapse time for an empty bubble collapsing without dissipation effects.
The plot distinctly illustrates the dependence of the maximum radius of the rebound bubble $R_1$ on the concentration of the ammonia solution, which increases from 19\% of $R_0$ for the least concentrated solution to 57\% of $R_0$ for the most concentrated solution.
On the other hand, $R_0$ remained nearly constant at 1.5 mm ($\pm$ 6\% due to laser power oscillation, see Table \ref{tab:radius}) independently of the ammonia concentration.
This pointed out that the influence of the ammonia content on the first bubble oscillation is negligible, suggesting that the physics of the plasma at generation is not altered by the presence of ammonia in solution, and that mass transfer effects do not affect the growth of the bubble.
\begin{table}
\caption{\label{tab:radius} Average radius of the bubbles at maximum expansion $R_0$, and average maximum radius of the first rebound $R_1$ for each ammonia concentration $w_{\text{NH}_3}$ exploited. $\sigma_{R_i}$ indicates the standard deviation of the measurements}
\begin{ruledtabular}
\begin{tabular}{lcccc}
$w_{\text{NH}_3}$ [-] &  $R_0$ [mm] & $\sigma_{R_0} \times 10^6$ & $R_1$ [mm] & $\sigma_{R_1} \times 10^6$\\
\hline
0 & 1.53 & 6.8 & 0.29 & 9.7\\
0.01 & 1.62 & 6.5 & 0.45 & 12.9\\
0.02 & 1.55 & 6.5 & 0.60 & 7.3\\
0.03 & 1.53 & 13.1 & 0.69 & 11.8\\
0.04 & 1.51 & 10.2 & 0.79 & 8.2\\
0.05 & 1.50 & 9.3 & 0.85 & 8.7\\
\end{tabular}
\end{ruledtabular}
\end{table}
Nevertheless, as shown in the nested plot in Figure \ref{fig:rt}(a), and in Figure \ref{fig:rt}(b), the bubble lifetime was gradually prolonged up to approximately 6\% with increasing ammonia concentration.
\par
\begin{figure*}
    \centering
    \includegraphics[width=0.98\textwidth]{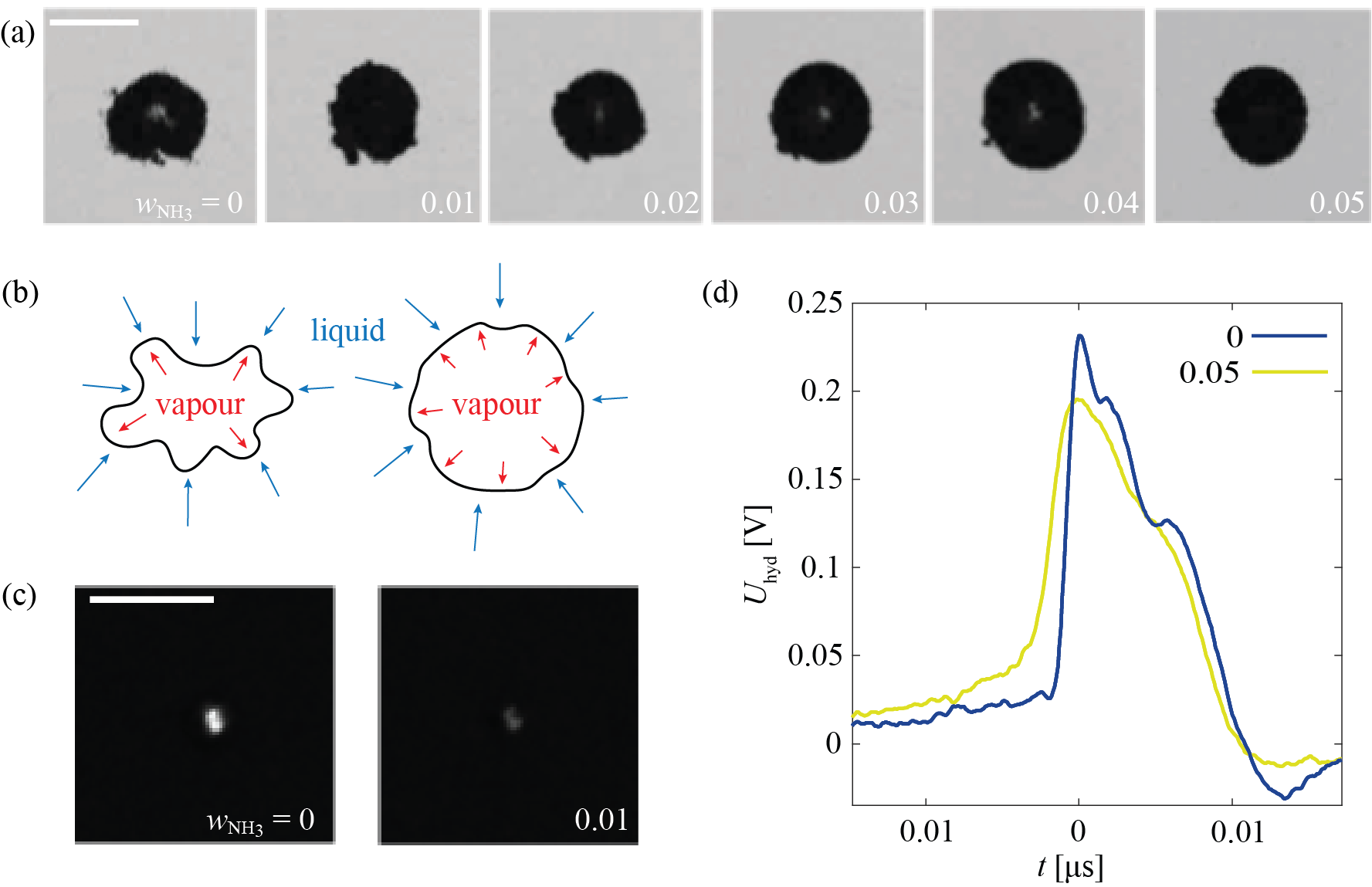}
    \caption{(a) Shadowgrams of the first rebound bubble for different ammonia mass fraction in water $w_{\ce{NH3}}$. The pictures display the rebound bubble taken at about the same expansion stage. The white line indicates 0.5-mm scale.
    (b) Illustration of the possible effect of vapor at the final stage of the bubble collapse. (c) Snapshots of the bubble in the final stage of the collapse (filmed at 10 million frames per second with an exposure time of 50 ns) in aqueous solutions with $w_{\text{NH}_3} = 0$ and $0.01$. The white line indicates the 0.5-mm scale.
    (d) Hydrophone signal over time recorded at collapse of two bubbles in aqueous ammonia at $w_{\ce{NH3}}$ of 0 and 0.05. The peak of the signals is centered at $t = 0$. }
    \label{fig:deformation}
\end{figure*}

 Furthermore, we noticed additional details strictly related to ammonia content in solution.
For instance, high-speed shadowgraphs, displayed in Figure \ref{fig:deformation}(a), show that the rebound bubble in pure water has a rather irregular shape with various defects at its interface.
The development of these deformations seems to be prevented by an augmentation of the ammonia content in solution as schematically represented in Figure \ref{fig:deformation}(b).
The bubble sketched on the left-hand side has a lower vapor content, leading to a larger growth of re-entrant jets and surface perturbations. On the contrary, greater vapor pressure in the bubble on the right-hand side opposes to the liquid contraction, leading to a more spherical collapse.
\par Observation of snapshots of the last instant of the collapsing bubble displayed in Figure \ref{fig:deformation}(c) reveal a brighter hot-spot for bubbles collapsing in pure water than in aqueous ammonia at $w_{\text{NH}_3} = 0.01$, indicating a larger energy density achieved due to a smaller minimum radius at collapse.
Although light emission is no longer detected at larger ammonia concentrations, the displayed snapshots of collapsing bubbles in the most diluted solutions show multiple irregular hot-spots.
\par The hydrophone signals $U$ over time, displayed in Figure \ref{fig:deformation}(d), also pointed out the dependence of the emitted shock waves on the ammonia content in solution.
The signal of the shock wave emitted upon bubble collapse in pure water ($w_{\ce{NH3}}=0$) exhibits multiple peaks, while it smooths out at larger ammonia concentrations ($w_{\ce{NH3}}=0.05$).
In addition, we surprisingly observed that the pressure build up recorded by the hydrophone before the main shock is more pronounced at larger ammonia concentration.

\subsection{Effects of the gaseous phase on the bubble dynamics}
We noticed several details about cavitation bubbles in aqueous ammonia at increasing concentration, namely (i) longer collapse time, followed by larger rebound bubbles, (ii) increased regularity of the spherical microscopic shape of rebound bubble, (iii) reduced light emission at collapse, (iv) enhanced regularity of the radiated shock's shape at collapse, and (v) larger liquid pressure build up prior to final collapse.
These observations, strictly related to the bubble contents, evidence an increased bubble internal pressure during the collapse, likely due to an increased vapor pressure of the two-component aqueous ammonia solution.
In particular, the longer collapse time, followed by larger rebounds, is likely due to the cushioning effect of the compressed gaseous phase, whose pressure increases with increasing ammonia content \cite{VazquezPutterman,Hopkins,TinguelyThesis}.
\par Bubble shape's irregularities, probably due to a loss of sphericity during the last moments of its collapse, decrease when the ammonia concentration is increased.
Although we are unable to observe the very final stage of the collapse, a faster increase of bubble pressure at collapse reduces the retraction of the liquid, increasing the minimum radius at collapse, and inhibits the crumpling of the bubble, thus the growth of deviations from sphericity, as illustrated in Figure \ref{fig:deformation}(b).
Such deviations from sphericity are likely brought on by the development of re-entrant non-piercing jets and/or hydrodynamic instabilities at the bubble-liquid interface such as Rayleigh-Taylor instabilities \cite{SupponenLuminescence,Ohl,Obreschkow2,HilgenfeldtLohseBrenner,BrennerLohseDupont,SinibaldiOcchiconePereira}. 
Optical systems, as the one employed for this work, are well known to generate bubbles with surface perturbations \cite{SupponenLuminescence,Baghdassarian}.
Our reasoning about an augmented bubble pressure at collapse is also backed by visualizations of the light emitted at bubble collapse.
\par Observations of light emission are consistent with the works of \citeauthor{Toegel} \cite{Toegel}, \citeauthor{VazquezPutterman} \cite{VazquezPutterman}, \citeauthor{Moss} \cite{Moss}, and \citeauthor{Hopkins} \cite{Hopkins}, who also observed higher light intensity with lower water vapor pressure within the bubble.
\par Finally, the analysis of the shock waves emitted at bubble collapse support the idea of a faster increase of the bubble internal pressure at collapse due to the increased vapor pressure in more concentrated ammonia solutions.
The indented shock wave signal recorded at bubble collapse in water, displayed in Figure \ref{fig:deformation}(d), probably arises due to successive collapses of different bubble fragments, whereas the smoother signal for the bubble in concentrated aqueous ammonia is synonym with more uniform bubble collapse.
Moreover, the more pronounced pressure build up in the liquid surrounding the bubbles in more concentrated aqueous ammonia is likely attributed to a greater deceleration of the liquid surrounding the bubble as the internal pressure of the latter increases \cite{Rayleigh,SupponenShockwaves}.
\par All else equal, i.e. $p_\infty$, $R_0$, and $\zeta$, a larger bubble internal pressure would only occur if a larger amount of matter in gaseous form is compressed during the collapse.
In this regard, although we cannot exclude that the non-condensable gas within the bubbles may have an effect on their dynamics \cite{Akhatov1}, we argue that our observations are not a consequence of the varying partial pressure of non-condensable components.
Specifically, non-condensable gases within laser-induced cavitation bubbles may result from vaporization of dissolved non-condensable gas at bubble generation, chemical reactions from plasma recombination, and diffusion from the liquid.
For bubbles similar to those presented in this work, diffusion effects were shown to be negligible with respect to the other two gas sources \cite{Plesset,Baghdassarian,BrennerHilgenfeldtLohse,Akhatov1}.
Furthermore, laser-generated gas due to chemical reactions is assumed to be proportional to the initial plasma energy density, hence to the bubble potential energy $E_\text{p0}$ \cite{Liang,SupponenLuminescence}, which was nearly constant.
We could also neglect any laser-generated gas due to the different liquid composition, as the dissociation energy of water and ammonia is similar (498 kJ/mol for the first \ce{O-H} bond in water and 435 kJ/mol for the \ce{NH2-H} bond of ammonia) \cite{Szwarc,Cottrell}.
In turn, the constant bubble volume at maximum expansion highlights the independence of the plasma nature on the ammonia content.
Conversely, this would likely reflect on the bubble size, absorbed energy, and/or energy partitioning at bubble generation.
This actually corroborates the effectiveness of employing aqueous ammonia as a successful tool for the investigation of cavitation bubbles.
Finally, vaporization due to plasma generation, proportional to the dissolved gas saturation, may give rise to variations of the partial pressure of non-condensable gas.
As mentioned before, \citeauthor{Narita} \cite{Narita} reported a decreased dissolved oxygen in aqueous ammonia with increasing ammonia concentration.
However, an increased ammonia content would translate into a decreased vaporized non-condensable gas, which contravene our observations of increased bubble pressure.
\par Altogether, despite we are unable to directly probe the bubble contents and their behaviour during the bubble collapse, our observations provide evidence that the vapor contained within a collapsing bubble is compressed.
As claimed in several numerical works present in literature \cite{Akhatov1,Fujikawa,Szeri,StoreySzeri,Magaletti,FusterHaukeDopazo,Bergamasco,FusterMontel,Aganin}, irrespective if equilibrium conditions are meet at the interface or not, condensation kinetics of vapors within the bubble are slower than the volume reduction rate.
Similar works have also been published recently. \cite{Ohashi,Kobayashi,Qin,Peng}
Vapors are then unable to fully condense during the collapse due to the thermal lag at the bubble interface, likely transitioning to the non-condensable gaseous state and providing the bubble with stronger means to resist and delay its collapse.
The development of a thermal boundary layer near the bubble wall was introduced earlier by \citeauthor{Fujikawa} \cite{Fujikawa}, \citeauthor{Yasui} \cite{Yasui}, \citeauthor{StoreySzeri} \cite{StoreySzeri}, and \citeauthor{Akhatov1} \cite{Akhatov1} in their numerical models.
In light of these theories, as the vapor pressure promptly rises with increasing ammonia concentration in the liquid (Figure \ref{fig:phaseDiagram}(a)), a greater amount of vapor remains trapped into the bubble at collapse due to the higher heat dissipation required for condensations at the bubble wall.
\par It is to be noted that in our experiments, the compressed gaseous phase is a two-component mixture of ammonia vapor and water vapor.
As shown in Figure \ref{fig:phaseDiagram}(a), the water vapor partial pressure decreases in more concentrated solutions, whereas the ammonia vapor partial pressure quickly increases.
Therefore, during the bubble collapse at increasingly concentrated aqueous ammonia solutions, the ratio of ammonia to water vapor that gets compressed rapidly changes.
Moreover, as the binodal curve of ammonia locates well above that of water (Figure \ref{fig:phaseDiagram}(b)), ammonia condenses more slowly and under higher pressure than water vapor.
It is however interesting to note that supercritical conditions may be reached more easily for ammonia.
Furthermore, in our multicomponent system, a mass-transfer boundary layer may also concurrently limit the condensation process along with heat transfer, adding an additional level of complexity to the problem.
\par In addition to supporting compression of vapors, it is to be noted that our results seem to be in contrast with theories focused on the idea that luminescence is due to shock-initiated thermal emission within the collapsing bubbles \cite{Moss,Yasui}.
In particular, \citeauthor{Evans} \cite{Evans} found that small deformations of a nearly spherical converging shock wave increase as the shock converges, eventually limiting the process of light emission.
This would be the case of our bubbles in more diluted solutions, which were observed with a more deformed surface.
Nevertheless, they emitted more light at collapse than the most spherical bubbles in more concentrated aqueous ammonia.
Our results are thus more in line with the theory of luminescence phenomena due to adiabatic compression of the bubble contents, also supported by \citeauthor{BrennerHilgenfeldtLohse} \cite{BrennerHilgenfeldtLohse}.
\par Finally, it is interesting and surprising to observe that the presence of ammonia did not influence the maximum radius of the bubbles.
This highlights the purely inertial nature of the bubble growth.
Furthermore, as the rebound ensuing the collapse is largely affected by the ammonia content, our results evidence that a higher concentration of ammonia in the water build the bubble contents up, likely increasing the vapor pressure within the bubble, highlighting the effectiveness of this technique to study the effect of the vapor pressure on cavitation bubbles.

\subsection{Energy dissipation at bubble collapse}
\par The energy balance in Equation \ref{eq:energybalance} revealed that a mere 1\% of the initial potential energy of the bubble $E_\text{p0}$ was reconverted into potential energy at the rebound $E_\text{p1}$ in pure water, resulting in a substantial normalized damping $\widehat{D}_\text{tot}$ of the bubble oscillation of approximately 99\%, where $\widehat{D}_\text{tot} = D_\text{tot}/E_\text{p0}$.
Conversely, in aqueous solutions with increasing ammonia content, $\widehat{D}_\text{tot}$ gradually decreases down to approximately 80\% at $w_{\ce{NH3}}=0.05$.
\par
\begin{figure*}
    \centering
    \includegraphics[width=0.75\textwidth]{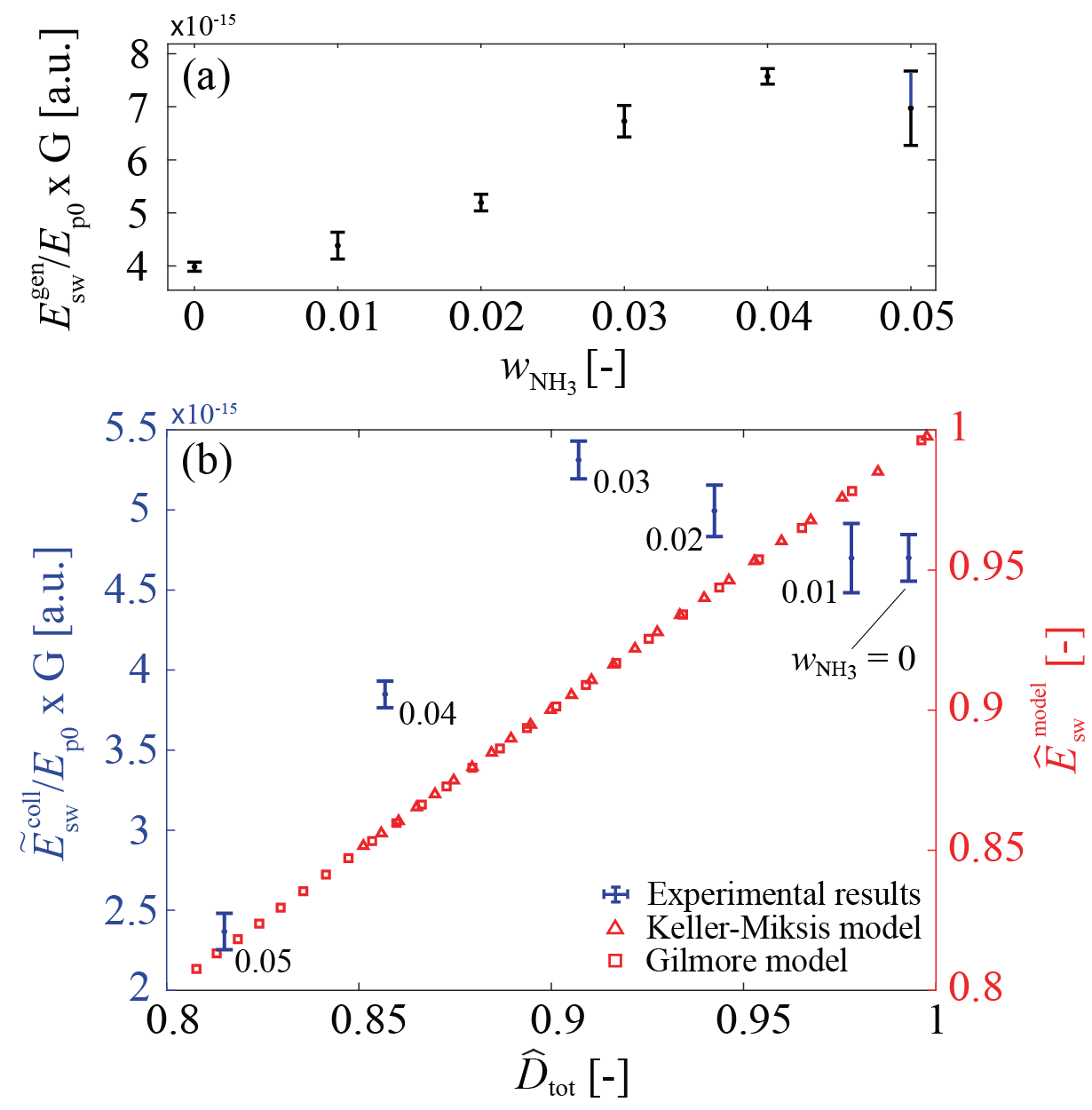}
    \caption{(a) Energy of shock waves emitted at bubble generation, normalized by the potential energy of the bubble at maximum expansion, as a function of $w_{\ce{NH3}}$. 
    (b) Energy of shock waves emitted at bubble collapse, normalized by the potential energy of the bubble at maximum expansion (experiments left y-axis, and simulations right y-axis), as a function of $\widehat{D}_\text{tot}$. 
    In both (a) and (b), points with error bars indicate the average normalized energy of the shock waves at bubble collapse obtained from 10 experiments in different aqueous ammonia with their relative standard deviation. 
    Triangles and squares in (b) refer to the energy of shock wave at bubble collapse normalized by the potential energy of the bubble at maximum expansion obtained from simulations with the \citeauthor{KellerMiksis} \cite{KellerMiksis} model and the \citeauthor{Gilmore} \cite{Gilmore} model, respectively, for bubbles at different initial non-condensable gas pressure $p_\text{g0}$.}
    \label{fig:sw}
\end{figure*}

 Besides bubble dynamics, we have investigated the shock waves emitted by the cavitation bubbles.
Figure \ref{fig:sw}(a) (left y-axis) displays the energy of the shock waves emitted at bubble generation $E_\text{sw}^\text{gen}$, normalized by the potential energy of the bubbles $E_\text{p0}$, as a function of the ammonia concentration in solution $w_{\text{NH}_3}$.
As mentioned in Section \ref{sec:methods}, the shock wave energy is multiplied by $G$ and reported in arbitrary units as a calibration of the hydrophone system was not essential for this work.
We observe an increase of $E_\text{sw}^\text{gen}$ with increasing $w_{\text{NH}_3}$.
Owing to the constant laser energy, hence initial plasma energy and bubble potential energy $E_\text{p0}$, we assume all shock waves at bubble generation in aqueous ammonia at different concentrations to be initially alike. 
Furthermore, as we do not notice a clear trend of $U_\text{max}^\text{gen}$, we assume the difference in $E_\text{sw}^\text{gen}$ only due to the influence of the diverse nature of the fluid.
From measurements of the shock waves energy at bubble generation, we then define a correction factor $\phi$ as
\begin{equation}
    \phi = \frac{\bar{E}_\text{sw}^\text{gen}-\bar{E}_\text{sw}^\text{gen}(w_{\text{NH}_3}=0)}{\bar{E}_\text{sw}^\text{gen}(w_{\text{NH}_3}=0)},
\end{equation}
where $\bar{E}_\text{sw}^\text{gen}$ is the average value of $E_\text{sw}^\text{gen}$ at some value of $w_{\text{NH}_3}$.
Such $\phi$ factor accounts for the overestimation of shock wave energy due to the presence of ammonia with respect to the case of water ($\phi = 0$ for $w_{\text{NH}_3}=0$).
Consequently, we compute a corrected shock wave energy at bubble collapse $\widetilde{E}_\text{sw}^\text{coll}$ as
\begin{equation}
    \widetilde{E}_\text{sw}^\text{coll} = E_\text{sw}^\text{coll} (1-\phi).
\end{equation}
The shock wave energy at bubble collapse $\widetilde{E}_\text{sw}^\text{coll}$ normalized by $E_\text{sw}^\text{gen}$, reported in Figure \ref{fig:sw}(b) (left y-axis), exhibited surprisingly a non-monotonic behaviour as a function of $\widehat{D}_\text{tot}$ (intrinsically related to $w_{\ce{NH3}}$ as shown in Figure \ref{fig:rt}(a)), peaking at 0.03 ammonia mass fraction, before eventually declining for more concentrated solutions.
It is to be noted that plotting as a function of $\widehat{D}_\text{tot}$ as in Figure \ref{fig:sw}(b) allows a direct comparison between experimental and numerical results albeit the bubble internal pressure is unknown for experiments.

\par This result seems to contradict the predictions of analytical models based on the \citeauthor{KellerMiksis} \cite{KellerMiksis} and \citeauthor{Gilmore} \cite{Gilmore} models, as shown in Figure \ref{fig:sw}(b) (right y-axis), where the value of the energy emitted in the shock wave at bubble collapse $\widehat{E}_\text{sw}^\text{model}$, where $\widehat{E}_\text{sw}^\text{model} = E_\text{sw}^\text{model}/E_\text{p0}$, for different values of initial non-condensable gas $p_\text{g0}$ is per construction comparable to $\widehat{D}_\text{tot}$ (see Appendix \ref{appAnalyticalModels}).
Interestingly, the trend of $\widehat{E}_\text{sw}^\text{model}$ is not reproduced by the reported experimental results.

\subsection{Mechanisms of energy dissipation}
The trend of $\widetilde{E}_\text{sw}^\text{coll}/E_\text{p0}$ shown in Figure \ref{fig:sw} indicates the presence of two regimes associated with bubble collapse in aqueous ammonia at various concentrations.
At low ammonia contents ($w_{\ce{NH3}}<0.03$), results indicate that the controlling mechanism of oscillation damping is different from that of liquid compressibility, as $\widetilde{E}_\text{sw}^\text{coll}/E_\text{p0}$ monotonically decreases as a function of $\widehat{D}_\text{tot}$, contrary to what we expected.
Conversely, at large ammonia concentrations ($w_{\ce{NH3}} \ge 0.03$), the liquid compressibility becomes the controlling mechanism of oscillation damping as $\widetilde{E}_\text{sw}^\text{coll}/E_\text{p0}$ is monotonically increasing with $\widehat{D}_\text{tot}$.
This experimental observation seems contrary to the predictions provided by simple models based on the \citeauthor{KellerMiksis} \cite{KellerMiksis} and \citeauthor{Gilmore} \cite{Gilmore} equation, where the main damping mechanism is controlled by liquid compressibility implying that the energy of the emitted wave and the damping of the bubble should be directly correlated. The reasons why simplified models based on \citeauthor{KellerMiksis} \cite{KellerMiksis} and \citeauthor{Gilmore} \cite{Gilmore} equation are not able to explain experimental observations should be topic of more careful future investigations. One possible explanation in light of diverse theories developed in the early 2000's is that deviations from sphericity are energy-consuming, which eventually dissipate energy into the liquid as heat. \cite{Brennen,Delale,GrandjeanThesis}
The bubble fragmentation and later coalescence phenomena have related extremely small length scales at which viscous dissipation may be magnified by several orders of magnitude compared to the spherical case. Although other mechanisms related to strongly nonlinear effects in the liquid or the presence of chemical reactions may also contribute to the effective damping of the bubble and will need detailed investigations, the results are in line with previous observations, as deviations from sphericity are the largest in the most diluted solutions, where we expect a smaller minimum radius at bubble collapse.
In this case, the damping effect due to the liquid compressibility seems to be overtaken by this second mechanisms of energy dissipation.
Our reasoning is further endorsed by the work of \citeauthor{Baghdassarian} \cite{Baghdassarian} and \citeauthor{SupponenLuminescence} \cite{SupponenLuminescence}, who reported that laser-induced bubbles are naturally formed with surface perturbations.
\par Finally, although this could be a mere coincidence, it is also interesting to note that the $E_\text{sw}^\text{coll}/E_\text{p0}$ curve peaks at approximately the same ammonia concentration whereby the ammonia vapor pressure overtakes the one of water (Figure \ref{fig:phaseDiagram}).
In practice, our results suggest that the perfectly spherical bubble collapse hypothesized in the \citeauthor{KellerMiksis} \cite{KellerMiksis} and the \citeauthor{Gilmore} \cite{Gilmore} models is inaccurate under certain conditions.
As these models account for dissipation mainly due to liquid compressibility, the use of these models may hinder the relevance of other dissipation effects during the collapse of the bubble which are not taken into account.
This may in turn compromise the accuracy of the bubble dynamics fitting from experimental data.

\section{Conclusion}
In this study, we provide experimental evidences about the effect of vapor content on transient cavitation bubbles:
\begin{itemize}
    \item Consistently with the predictions for single bubble sonoluminescence experiments, the vapor is trapped inside the bubble and compressed during the violent collapse. 
    Through an increase of the vapor pressure within the bubble by adding ammonia to water, we observe a striking role of the vapor on the amplitude of the rebound bubble.
    We infer that the progressively decreasing bubble deformation as a function of the ammonia concentration is due to a larger resistance of the bubble content to compression, which limits the growth of deformations and/or hydrodynamic instabilities as a consequence of the smaller values of the acceleration during the last stages of the collapse.
    \item The results are endorsed by the luminescence phenomena, whose brighter emission in cavitation bubbles generated in pure water, thus with a smaller minimum radius, is synonym of larger  energy density.
    \item The liquid pressure build up prior to the final collapse and the shape of the shock wave emitted successively also indicate that the bubble internal pressure increases faster during the bubble collapse at the instant of collapse for increasing ammonia in solutions.
    Our reasoning behind the compression of vapors is that vapor-liquid phase transition are restrained, likely by mass and heat transfer.
\end{itemize}
Moreover, we demonstrate that: 
\begin{itemize}
    \item While the initial growth and subsequent compression of a laser-induced bubble is nearly insensitive to the partial vapor pressure of the host fluid, the damping mechanisms controlling the amplitude of the rebound are greatly influenced by the vapor content.
    Liquid compressibility effects alone cannot explain the oscillation damping of collapsing bubbles. 
    Some phenomena that could explain the experimental observations include phase change effects during the last stages of the bubble collapse, chemical dissociation reactions and bubble fragmentation due to Rayleigh-Taylor instabilities, which may eventually lead to the appearance of extremely small scale agitation that would ultimately increase the viscous dissipation.
    \item These other dissipation phenomena may play a preponderant role particularly in bubbles with low internal pressure at collapse, hence diluted solutions, prevalent in most of real-life applications.
\end{itemize}

It would be however interesting to obtain absolute pressure measurement obtaining the calibration constant $G$ as a function of the ammonia content in solution.
Furthermore, it would be of great relevance to study the shock waves at bubble collapse in more details, ideally investigating experimentally the shock waves behaviour in the near-field.
\par Our findings may help improving the development of more sophisticated numerical models for improved prediction of the cavitation process.
We therefore point out the urge to include more precise phase transition prediction in numerical models and call into question the disregard of aspherical perturbed collapsing bubbles.
These findings also contribute to our general understanding of cavitation bubbles, with several opportunities for practical applications, including understanding and controlling cavitation erosion processes, regulate the collapse intensity for controlled sonochemical reactions, and improve cavitation-based biomedical technologies.
Moreover, they also provide novel information of cavitation in aqueous ammonia, whose occurrence is to be reckoned.
Alongside an already-extensive use in chemical plants, ammonia as carbon-free fuel and green hydrogen carrier has gained significant attention over the last decade \cite{Valera,Yapicioglu}.

\begin{acknowledgments}
This work was supported by the MSCA-ITN of the EU Horizon 2020 Research and Innovation program (Grant Agreement No. 813766), and the Swiss National Science Foundation (Grant No. 179018).
\end{acknowledgments}

\section*{Conflict of Interest}
The authors have no conflicts to disclose.

\section*{Author Contributions}
\textbf{Davide Bernardo Preso}: Conceptualization; Formal analysis; Investigation; Methodology; Validation; Visualization; Writing - original draft.
\textbf{Daniel Fuster}: Conceptualization; Formal analysis; Methodology; Supervision; Writing - review \& editing.
\textbf{Armand Baptiste Sieber}: Investigation; Methodology; Visualization; Writing - review \& editing.
\textbf{Danail Obreschkow}: Formal analysis; Methodology; Validation; Writing - review \& editing.
\textbf{Mohamed Farhat}: Conceptualization; Formal analysis; Funding acquisition; Project administration; Supervision; Writing - review \& editing.

\section*{Data Availability Statement}
The data that support the findings of this study are available from the corresponding author upon reasonable request.

\appendix

\section{Analytical models for bubble dynamics}
\label{appAnalyticalModels}
\par
\begin{figure}
    \centering
    \includegraphics[width=0.7\textwidth]{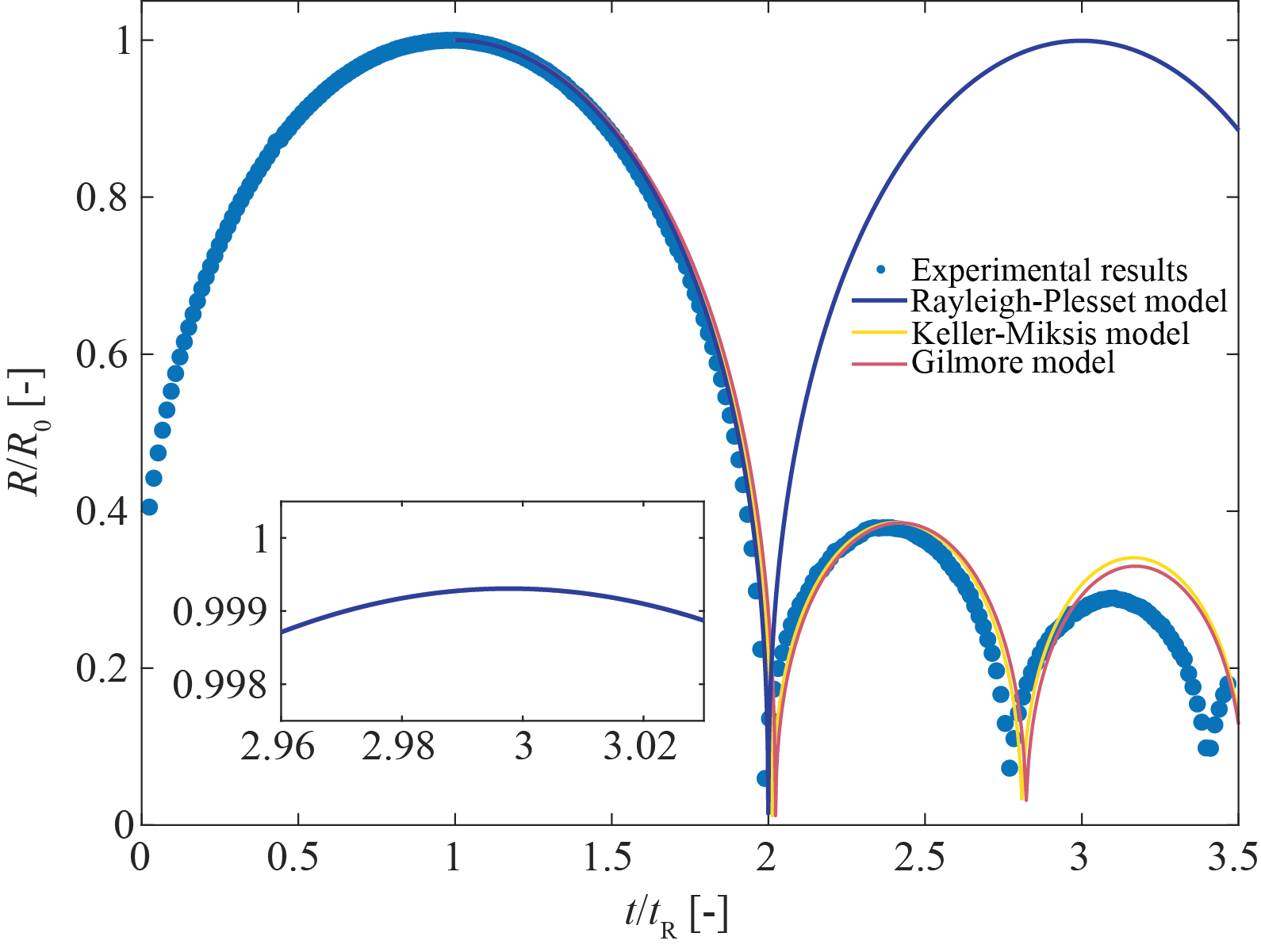}
    \caption{Radial evolution in time, in normalized coordinates, of cavitation bubbles obtained from experiments and numerical models. In both cases, $R_0$ is 1.5 mm. The $p_\text{g0}$ used in the simulations is 50 Pa, 48 Pa, and 42 Pa, for the Rayleigh-\citeauthor{Plesset} \cite{Plesset}, \citeauthor{KellerMiksis} \cite{KellerMiksis}, and \citeauthor{Gilmore} \cite{Gilmore} models, respectively.}
    \label{fig:figappendix}
\end{figure}

 The Rayleigh-\citeauthor{Plesset} \cite{Plesset}, the \citeauthor{KellerMiksis} \cite{KellerMiksis}, and the \citeauthor{Gilmore} \cite{Gilmore} models are numerically solved with a fourth-order Runge-Kutta method with adaptive time stepping.
With the last two, the internal pressure of a collapsing bubble is estimated by fitting the experimental data up to the second oscillation.
The model is constrained to find the value of $p_\text{g0}$, the free parameter, with which the model best fits the maximum amplitude of the rebound bubble after collapse.
The parameters used for simulations with the \citeauthor{KellerMiksis} \cite{KellerMiksis} and the \citeauthor{Gilmore} \cite{Gilmore} model in this work are reported in Table \ref{tab:KMG}.
An example of fitting with the two models is reported in Figure \ref{fig:figappendix}.
In the \citeauthor{KellerMiksis} \cite{KellerMiksis} and the \citeauthor{Gilmore} \cite{Gilmore} models, we neglected viscous dissipation effects for simplicity.

\begin{table}
\caption{\label{tab:KMG}Physical data used for numerical simulations with the \citeauthor{KellerMiksis} \cite{KellerMiksis} and the \citeauthor{Gilmore} \cite{Gilmore} models}
\begin{ruledtabular}
\begin{tabular}{ccccc}
$p_\text{v}$ (@\SI{20}{\celsius}) [Pa]  &  $\rho^\text{mix}$ [kg m$^{-3}$]  &  $c$ [m s$^{-1}$] & $\gamma$ & $p_\infty$ [kPa]\\
\hline
2502.8 & 998.2 & 1479 & 1.35 & 97.3\\
\end{tabular}
\end{ruledtabular}
\end{table}

\bibliography{aipsamp}

\end{document}